\documentstyle[aps,prl,twocolumn]{revtex}
\begin{document}
\twocolumn[\hsize\textwidth\columnwidth\hsize
\csname@twocolumnfalse\endcsname
\draft
\title{Controlled Drift of Indirect Excitons in Coupled Quantum
Wells: Toward Bose Condensation in a Two-Dimensional Trap}
\author{V. Negoita and D.W. Snoke}
\address{Department of Physics and Astronomy, University of
Pittsburgh\\ 3941 O'Hara St.\\ Pittsburgh, PA 15260, USA}
\author{K. Eberl}
\address{Max-Planck-Institut f\"ur Festk\"orperforschung\\
Heisenbergstr. 1\\ 70506 Stuttgart, Germany}
%

%\date{August 27, 1998}
\maketitle

%\newpage

\begin{abstract}
We have succeeded in trapping indirect excitons in coupled quantum
wells in a harmonic potential minimum via inhomogeneous applied
stress and electric field. These excitons exhibit a strong Stark
shift (over 60 meV), long lifetime (100 ns), and high diffusivity
(1000 cm$^2$/s). This approach is very promising for obtaining the
high exciton density needed for Bose condensation of excitons in two
dimensions.
\end{abstract}
]
\pacs{71.35.Lk; 73.20.Dx; 78.66.-w}
\narrowtext

Over the past 15 years, several experiments have indicated evidence
of Bose effects or Bose condensation of excitons in semiconductors
\cite{timofeev,pegh,snokeev,snoke-lin,wolfe-lin,naga,fuku,abstreiter1,abstreiter2%
,fortin,goto1,goto2,wolfe-kim}. These experiments mostly fall into
four basic categories: evidence based on spectral lineshape analysis
following incoherent generation of the excitons, which shows
narrowing of the exciton luminescence lines at high densities
\cite{timofeev,pegh,snoke-lin,wolfe-lin,fuku}, evidence based on
comparison of total luminescence intensities of two different
excitonic species, by which the relative populations can be deduced
\cite{snokeev,wolfe-kim}, evidence based on light emission following
coherent generation of excitons in the ground state, which shows that
excitons remain in regions of phase space near the ground state for
time periods long compared to the exciton scattering time
\cite{naga,goto1,goto2}, and measurements of the transport of the
excitons which show fast expansion out of the creation region
\cite{snokeev,abstreiter2,fortin}. This body of evidence, while
important, lacks the dramatic ``smoking gun'' that has been seen in
alkali atoms in magneto-optical traps
\cite{cornell,ketterle}, namely, a {\em spatial} condensation into a
two-component distribution, a clear prediction of the theory of the
weakly interacting Bose gas which has no classical analog. It has
long been known \cite{trauer} that excitons in a harmonic potential
will also show this behavior if they undergo Bose condensation; a
method of creating a harmonic potential for excitons in bulk
semiconductors is well established \cite{trauer}, but so far,
experimental attempts with bulk semiconductors have not succeeded in
creating a density of excitons high enough for Bose condensation in
this kind of trap. 

Much recent attention has been given to indirect, or ``dipole,''
excitons in two-dimensional heterostructures
\cite{fuku,abstreiter1,abstreiter2,peeters,kash,rutgers,gilliland}. 
This system is appealing because (1) the excitons can have long
lifetimes due to the spatial separation of the electron and hole, (2)
the interaction between the dipole-aligned excitons is strongly
repulsive, so that crossover to a Fermi liquid state is not expected
at high density, and (3) the quality of semiconductor
heterostructures has been steadily increasing, so that true
two-dimensional physics can be studied. In a two-dimensional system,
Bose-Einstein condensation is not expected, but rather a
Kosterlitz-Thouless transition to a superfluid state
\cite{huang}, although J. Fern\'andez-Rossier, C. Tejedor, and R.
Merlin
\cite{merlin} have recently argued that the coupling of the excitons
to the photon states will allow them to undergo Bose condensation in
two dimensions.

Early experiments with this type of structure \cite{fuku} showed
evidence for Bose effects, but later work \cite{kash} showed that
localization due to random variations in the structures significantly
complicated the analysis of the luminescence lineshape. Recent
studies of similar stuctures \cite{abstreiter1,abstreiter2} have
shown quite promising results, including evidence for increased
diffusion out of the excitation region at high density and low
temperature. Other recent measurements of the diffusion of indirect
excitons have also shown fast expansion at high densities
\cite{gilliland}. Enhanced diffusion is expected for superfluid
excitons, but can also be attributed to other, classical effects
which also occur at high density, such as phonon wind
\cite{tikhodeev}.

In order to overcome the complications of localization and classical,
pressure-driven expansion,  X.J. Zhu, P.B. Littlewood, and T.M. Rice
\cite{rutgers} proposed a variation in which {\em inward pressure} on
the excitons is produced which confines them to a potential minimum.
As Nozier\'es \cite{noz} and others have pointed out, if a potential
minimum exists, true Bose condensation can occur in two dimensions
instead of a Kosterlitz-Thouless transition. Zhu, Littlewood, and
Rice envisioned that a potential minimum could be created by a
variation in the quantum well thickness. In this Letter, we report
the experimental accomplishment of a potential minimum for indirect
excitons in a two-dimensional plane via a different means. This
method creates a harmonic potential minimum for the excitons, so that
the telltale two-component spatial signature of Bose condensation can
occur, and it allows us to vary the depth of the potential minimum
via an external control.

The samples we use are GaAsAl$_x$Ga$_{1-x}$As coupled quantum well
structures fabricated via molecular-beam epitaxy (MBE) at the
Max-Planck-Instutute in Stuttgart; the substrate is heavily p-doped
and the capping layer is heavily n-doped in order to allow electric
field perpendicular to plane of the quantum wells. Fig. 1(a)
illustrates the band structure when electric field is applied; as
seen in Fig. 1(b), as the electric field is increased, the energy of
the indirect excitons undergoes a strong Stark shift to lower energy,
as also seen in previous studies (e.g., Refs.
\cite{stark1,stark2}.)  The spatial separation of electron and hole
into two separate planes also increases the lifetime of the excitons;
in our samples we measure lifetimes of the indirect excitons of
around 100 ns. 

We create a potential minimum for the excitons via externally
applied, inhomogeneous stress and electric field. Fig. 2 shows the
experimental geometry. The quantum well sample is clamped between two
metal plates, each with a small hole, and a pin is pressed against
the GaAs substrate, which has been polished on both surfaces prior to
the MBE fabrication. The pin creates a shear strain maximum in the
quantum wells, as well as a slight hydrostatic expansion; both of
these strain effects lead to an energy minumum for the excitons via
the Pikus and Bir deformation Hamiltonian
\cite{PB-note}, similar to the way in which inhomogeneous strain
leads to a potential energy minimum for carriers in bulk
semiconductors
\cite{trauer,wolfe-ge,wolfe-si}. Too much stress from the pin will
cleave the sample, of course, but springs on the back of the sample
help to prevent this, allowing a reproducible, controllable stress.

In addition, the pin is held at a fixed, negative voltage while the
clamping plates are connected to ground. This causes a current to
flow through the heavily-doped substrate, so that the voltage across
the quantum wells drops to zero far away from the pin. As seen in
Fig. 1, higher electric field corresponds to lower energy for the
indirect excitons, so that this effect also contributes to a
potential energy minimum for the excitons below the pin.

The entire assembly is placed in liquid or gaseous helium, and the
quantum wells are excited by a laser through the window of an optical
cryostat by means of a prism attached to the lower metal plate. The
force on the pin is controlled by a micrometer at the top of the
cryostat, as in Ref. \cite{wolfe-si}. Fig. 3 shows time-integrated 
luminescence from a coupled quantum well sample with 60 \AA~ GaAs
wells and 42 \AA~ Al$_{.3}$Ga$_{.7}$As barrier, taken with a CCD
camera on the back of an imaging spectrometer as the laser spot is
scanned across the surface of the sample. As seen in Fig. 3(a), a well
depth of more than 10 meV can be created, compared to the
inhomogeneous broadening in these samples of slightly less than 1
meV. When the voltage applied to the pin is set to zero, the same
time-integrated scan gives Fig. 3(b), which shows that the effect of
the variation in voltage is about the same as the effect of the shear
strain maximum. 

The fact that a potential minimum occurs is strongly connected to the
geometry which leaves the lower surface of the sample unconstrained. 
When the sample is placed on a glass slide, a potential energy {\em
maximum} is seen, since in this case the sample is compressed, and
the positive shift in energy due to the hydrostatic deformation
potential overwhelms the negative shift due to shear strain. We have
solved the static field equations for strain in the sample via
finite-element analysis, and the shifts in energy in both cases agree
with our calculations \cite{inpress}.

In a harmonic potential minimum in two dimensions, the critical
number for Bose condensation is given by
\begin{eqnarray} N_c & = &\sum_n n \frac{1}{e^{n\hbar\omega/k_BT}
-1}\\ \nonumber & = & \frac{(k_BT)^2}{(\hbar\omega_0)^2}\int \epsilon
d\epsilon \
\frac{1}{e^{\epsilon} - 1}\\ \nonumber & = & 1.8
\frac{(k_BT)^2}{(\hbar\omega_0)^2},
\end{eqnarray}
 where
$\omega_0 = (\alpha/m)^{1/2}$.  The shape of the well shown in Fig.
3(a) corresponds to a force constant of $\alpha = 65$ meV/mm$^2$,
approximating $U= \alpha x^2/2$ in the center of the trap. For a
temperature of 2 Kelvin and exciton mass on the order of the electron
mass, this critical number is approximately $10^7$. By comparison, a
single laser pulse from our dye laser contains more than $10^{11}$
photons. We have not seen evidence for Bose effects in this well,
however. Because we excite at $\lambda = 660$ nm, the excess energy
of the generated carriers is quite high, so that the carrier
temperature is well above 100 K for most of their lifetime, as
determined by fits to the band-edge luminescence from the substrate
at the same times, even when the sample is immersed in liquid helium.
At this temperature, the critical number is four orders of magnitude
higher. 

 The diffusion length of the excitons is also too short for
thermalization in the well at this temperature. Fig. 4 shows the
spatial profile of the indirect exciton luminescence at various times
after a laser pulse has created them about 400
$\mu$m from the center of the well. The expansion at early time
corresponds to a diffusion constant of over 1000 cm$^2$/s, similar to
that of Ref. \cite{gilliland}. (This fast expansion at early time is
essentially the same even with zero applied stress.) At late times,
as the exciton density drops, the expansion of the excitons slows
down, although the effect of drift due to the gradient in potential
energy is clearly seen.

For a lifetime of 100 ns and D = 1000 cm$^2$/s, the diffusion length
of the excitons is around 100 $\mu$m. By comparison, the equilibrium
spatial width of a classical gas in a harmonic potential well with 
$\alpha = 65$ meV/mm$^2$, determined approximately by the condition
$\alpha x^2/2 = 3k_BT/2$ \cite{trauer}, is over 500
$\mu$m.

As the exciton gas gets colder, the equilibrium spatial width
should become smaller, to less than 100 $\mu$m at 2 K. This also
indicates the importance of lower effective exciton temperature. We
believe that by creating the excitons with lower energy via
near-resonant excitation, we can significantly reduce the exciton
temperature in the future.

Another approach which may also aid the approach to Bose condensation
of excitons in this geometry may be adding a strong magnetic field,
which will reduce the spin degeneracy of the excitons, forcing higher
numbers of particles into few states, and which will also create a
more strongly repulsive interaction between the excitons.  Several
authors \cite{liberman,paquet,lerner} have argued that magnetic field
will enhance Bose effects of excitons; experimentally, Ref.
\cite{abstreiter2} reported a sharp increase of diffusivity of
indirect excitons above a critical threshold of magnetic field.

Recently, several authors \cite{merlin,sham,laik} have proposed
optical tests for the phase coherence which should appear in the
excitonic Bose condensate. Underlying all these approaches is the
fact that Bose condensation implies spontaneous phase coherence, and
since excitons couple to photon states, this phase coherence should
transfer to the photons, even in the absence of lasing. The method
proposed here of confining an exciton condensate to a trap is much
more amenable to these kinds of tests than methods which allow free
expansion of the exciton gas, since the ground state in this case is
well defined.

Finally, we note that the method we have used here to trap the
excitons may have other applications. Since excitons are charge
neutral, they do not respond to electric field, and it is therefore
difficult to control their motion. We have shown that the motion of
excitons in heterostructures can be controlled over  distances up to
100 $\mu$m via both inhomogeneous shear stress and inhomogeneous
electric field. In particular, variation of the voltage across the
quantum wells can be accomplished by depositing resistive patterns on
the surface via photolithography. Small ``wires'' for excitons can
therefore be created which carry excitons from place to place in
response to electric fields.

{\bf Acknowledgements}. This work has been supported by the National
Science Foundation as part of Early Career award DMR-97-22239. One of
the authors (D.S.) is a Cottrell Scholar of the Research Corporation.
We thank I. Hancu for early contributions to these experiments, and
L.M. Smith for helpful conversations.

\begin{figure}[t]
\caption{(a) Band structure of the coupled quantum well structures
used in this experiment. Indirect excitons are formed from electrons
in the lowest conduction subband and the highest valence subband.
(b) Peak photon energy of the two
luminescence lines from the structure, as a function of reverse bias voltage. Solid
circles: indirect (interwell) excitons, open circles: direct (intrawell) excitons.
The intensity of the intrawell excitons relative to the interwell excitons first
falls, then rises as the electric field is increased.}
\label{fig1}
\end{figure}
\begin{figure}[t]
\caption{Exerimental geometry for applying inhomogeneous strain and
electric field to the samples. The pin is pressed against the sample
with approximately 5 lb. of force.}
\label{fig2}
\end{figure}
\begin{figure}[t]
\caption{(a) Time-integrated image through an imaging spectrometer of
the indirect exciton luminescence as the laser spot is scanned across
the surface of the quantum well sample with 43 kV/cm applied field.
The point of lowest energy corresponds to the point directly below
the tip of the pin shown in Fig. 2. (b) Time-integrated image for the
same conditions but zero applied field. As seen in this comparison,
the inhomogeneous electric field gives a contribution to the trap as
large as that of the applied shear stress.}
\label{fig3}
\end{figure}
\begin{figure}[t]
\caption{Spatial profiles of the luminescence from the coupled
quantum well structure at various times after a short (5 ps) laser
pulse. The laser is focused about 400 $\mu$m from the center of the
trap shown in Fig. 3. As seen in this figure, the indirect excitons
drift toward the center of the well due to the gradient in potential
energy. The expansion at early time, before 50 ns, corresponds to a
diffusion constant of 1000 cm$^2$/s.}
\label{fig4}
\end{figure}
\end{document}